\documentclass[9pt,twocolumn,twoside]{opticajnl}
\journal{opticajournal} 

\setboolean{shortarticle}{true}

\usepackage{lineno}

\makeatletter
\providecommand{\onecolumngrid}{\clearpage\onecolumn}

\makeatother

\usepackage{amsmath,mathrsfs,amssymb,amsfonts,amsthm,mathtools}
\usepackage{graphicx,color}
\usepackage{physics}
\usepackage[T1]{fontenc}
\usepackage{empheq}
\usepackage{bm}
\usepackage{hyperref}
\hypersetup{
    colorlinks=true,
    linkcolor=blue,
    citecolor=blue,
    urlcolor=blue
}

\newcommand{\nn}{\nonumber}

\title{The Janus State: A Universal Lower Bound for Second-Order Coherence}
\author[1,2]{Arash Azizi}

\affil[1]{The Institute for Quantum Science and Engineering, Texas A\&M University, College Station, TX 77843, U.S.A.}
\affil[2]{Department of Physics and Astronomy, Texas A\&M University, College Station, TX 77843, U.S.A.}

\affil[*]{sazizi@tamu.edu}

\begin{abstract}
A single-mode squeezed vacuum is a foundational quantum state that, despite its nonclassical nature, exhibits classical-like, super-Poissonian photon statistics. This feature motivates a ``quantum-of-quantum'' inquiry: can the superposition of two such states generate the opposite behavior—strongly sub-Poissonian light? We demonstrate that the ``Janus state,'' a coherent superposition of two squeezed vacua with opposing orientations, achieves precisely this. Our exact analytic solution reveals a \textbf{universal lower bound} on second-order coherence, showing that $g^{(2)}$ cannot be driven below $1/2$. The mechanism is tuned interference that suppresses two-photon events. Beyond this asymptotic bound, we identify a practical minimum of $g^{(2)} \approx 0.567$ at moderate squeezing, defining an accessible ``sweet spot.'' While requiring a minimal non-Gaussian element for its creation, the Janus state establishes a definitive performance limit for engineering sub-Poissonian photon statistics from Gaussian resources, with a clear path toward quantum applications.
\end{abstract}

\begin{document}

\maketitle

\section{Introduction}
Nonclassical states of light are central to secure communication, precision metrology, and scalable photonic information processing \cite{braunstein2005quantum,weedbrook2012gaussian,kimble2008quantum,Agarwal2012book}. A widely used diagnostic is the normally ordered second–order coherence $g^{(2)}$, whose reduction below unity signals sub-Poissonian statistics within Glauber’s coherence theory \cite{glauber1963coherent, sudarshan1963equivalence,    Mandel_Wolf1995}. Conventional routes to sub-Poissonian statistics typically rely on heralding, strong optical nonlinearities, or interference with sizable coherent displacements \cite{slusher1985observatio,Ourjoumtsev2007,Parigi2007}.

Meanwhile, squeezed-vacuum sources have matured into robust platforms, motivating design rules that leverage only Gaussian “primitives” \cite{Schleich_Wheeler1987, Schleich1988PRA, vahlbruch2016detection,Park2024}. A key feature is that a single-mode squeezed vacuum—already a quantum state—exhibits classical-like, super-Poissonian photon statistics, $g^{(2)}=3+1/\sinh^{2}r>3$ \cite{Loudon2000quantum}. This sets up a sharper “quantum-of-quantum’’ question: can a coherent superposition of two such states, tuned only by relative phase and orientation, overturn the super-Poissonian regime of its constituents and produce strongly sub-Poissonian statistics?

We christen this superposition the \emph{Janus state}, after the two-faced Roman god of duality, to emphasize its construction from squeezed-vacuum components with opposing phase-space orientations. Superpositions of squeezed states have been explored for decades \cite{Sanders1989Superposition, Buzek1992Superpositions,  Obada1997}; adopting a concise moniker—akin to the “cat state’’—is pragmatic and highlights an interference property with operational utility.

A compact intuition follows from the small-squeezing expansion. For $r\ll1$, $\ket{\xi(\theta)} \approx \ket{0}-\frac{e^{i\theta}}{\sqrt{2}}\,r\,\ket{2}$
up to $\mathcal{O}(r^{2})$. Taking two squeezers with equal strength but opposite orientations and forming the subtraction,
\[
\ket{\xi(\theta)}-\ket{\xi(\theta+\pi)} \;\propto\; \ket{2},
\]
the vacuum component cancels while the two-photon amplitude adds. Since $g^{(2)}(\ket{n})=1-\tfrac{1}{n}$, this picture predicts the asymptotic limit $g^{(2)}\!\to\!g^{(2)}(\ket{2})=\tfrac{1}{2}$.

Building on this intuition, we provide a closed-form analysis of the Janus state and establish a platform-independent lower boundary on second–order coherence:
\begin{equation}
\lim_{r \to 0} g^{(2)}_{\min}(r) = \tfrac{1}{2}. \label{eq:half_bound}
\end{equation}
Moreover, we identify a practical local minimum $g^{(2)} \approx 0.567$ at a moderate squeezing $r \approx 0.34$, defining an experimentally accessible “sweet spot.”

\begin{figure}[t]
    \centering
    \includegraphics[width=\columnwidth]{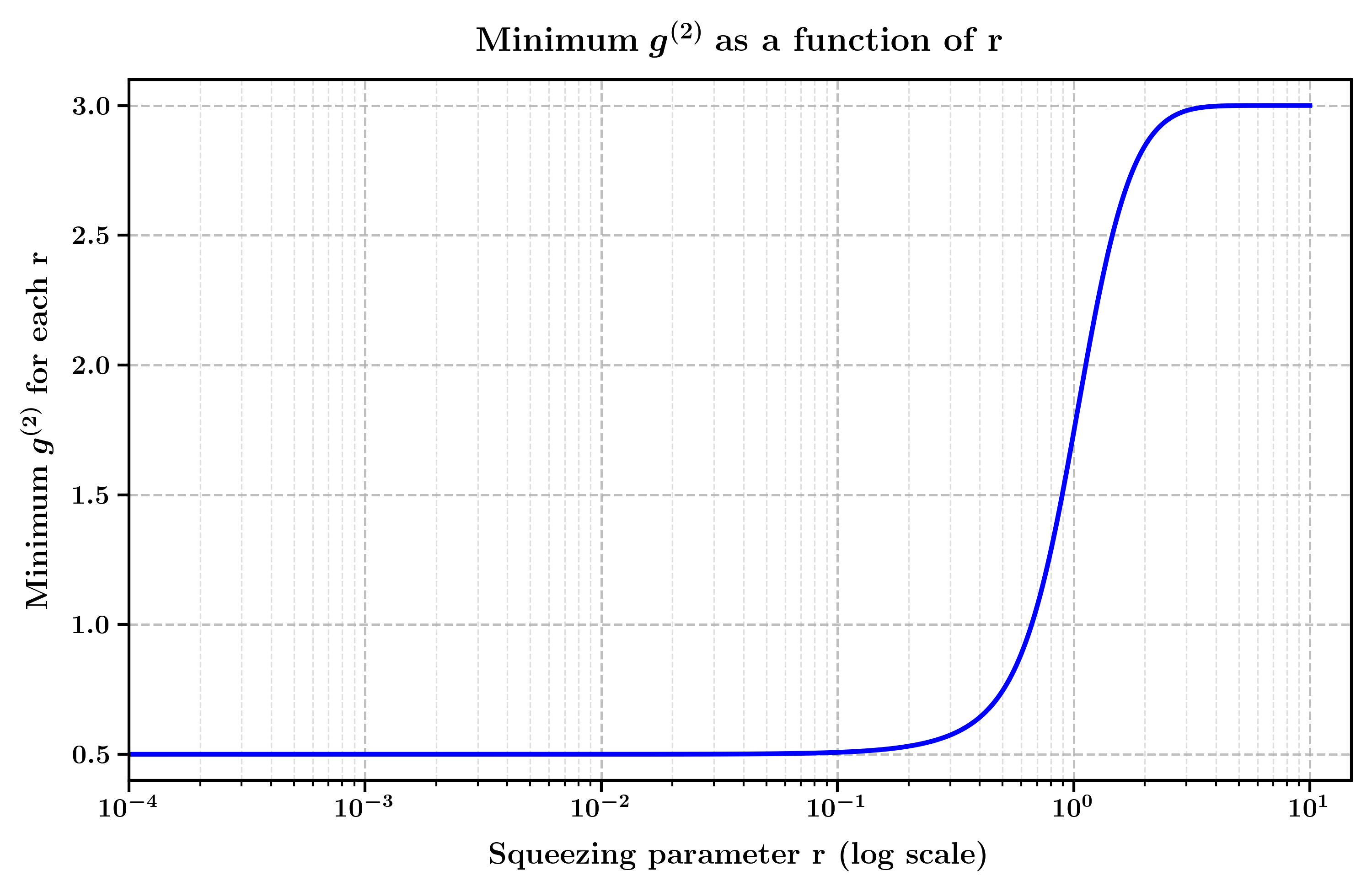}
    \caption{%
        \textbf{Global minimum of $g^{(2)}$ vs.~$r$.}
        The universal lower bound $g^{(2)}\!=\!1/2$ is approached as $r\!\to\!0$;
        a practical local minimum $\approx 0.567$ occurs near $r\!\approx\!0.34$.
    }
    \label{fig:g2_global_scan}
\end{figure}

The implications extend beyond $g^{(2)}(0)$. The same interference logic enables controlled tuning of higher-order correlations and non-Gaussian features—including Wigner-function negativity as a bona fide resource \cite{Kenfack2004,Albarelli2018Resource}. This places Janus states as compact resources for integrated photonics and hybrid continuous/discrete-variable protocols \cite{Menicucci2006,Andersen2015Hybrid}, and connects naturally to quantum-enhanced sensing where tailored photon statistics boost Fisher information \cite{Giovannetti2006,Pezzesmerzi2018RMP}. In what follows, we derive the general expressions, identify the optimal parameters, and outline a realistic path to experimental preparation. Figure~\ref{fig:g2_global_scan} summarizes the two key features—the $1/2$ boundary and the $r\!\approx\!0.34$ sweet spot.

\section{The Janus state and its photon statistics}

We define the \emph{Janus} state as a coherent superposition of two single-mode squeezed vacua,
$|\xi\rangle \equiv |r e^{i\theta}\rangle$ and $|\zeta\rangle \equiv |s e^{i\phi}\rangle$, with relative quadrature orientation
$\Delta \equiv \theta-\phi$:
\begin{equation}
|\psi\rangle \;=\; |\chi|\,|\xi\rangle \;+\; |\eta|\,e^{i\delta}\,|\zeta\rangle ,
\label{eq:superposition_main}
\end{equation}
where $|\chi|,|\eta|\in\mathbb{R}_{\ge 0}$ are amplitudes and $\delta$ is the superposition phase.
Photon statistics are characterized by the normally ordered second-order coherence
$g^{(2)} \equiv \langle a^{\dagger 2} a^{2}\rangle/\langle a^{\dagger} a\rangle^{2}$.
For a \emph{single} squeezed vacuum one has
$g^{(2)} = 3 + 1/\sinh^{2}\!r > 3$ (strong super-Poissonian regime) \cite{Loudon2000quantum}.

For compactness, set
\[
x\equiv\tanh^{2}\!r,\qquad
y\equiv\tanh^{2}\!s,\qquad
z\equiv\tanh r\,\tanh s\,e^{i\Delta}.
\]
With these definitions, write
\begin{align}
\mathcal{N}
&= |\chi|^{2}\,\frac{x(2x+1)}{(1-x)^{2}}
 \;+\; |\eta|^{2}\,\frac{y(2y+1)}{(1-y)^{2}} \nonumber\\
&\quad + 2\,\mathscr{Re}\!\left[
\chi\,\eta^{*}\;
\frac{(1-x)^{1/4}(1-y)^{1/4}\,z(2z+1)}{(1-z)^{5/2}}
\right],
\label{eq:N_def}\\[4pt]
\mathcal{D}
&= |\chi|^{2}\,\frac{x}{1-x}
 \;+\; |\eta|^{2}\,\frac{y}{1-y} \nonumber\\
&\quad + 2\,\mathscr{Re}\!\left[
\chi\,\eta^{*}\;
\frac{(1-x)^{1/4}(1-y)^{1/4}\,z}{(1-z)^{3/2}}
\right].
\label{eq:D_def}
\end{align}
The second-order coherence then takes the compact form
\begin{equation}
g^{(2)}(\psi) \;=\; \frac{\mathcal{N}}{\mathcal{D}^{2}},
\end{equation}
with the state normalized by
\begin{align}
|\chi|^{2}+&|\eta|^{2} \label{eq:psi_norm_general}\\
+& 2\,\mathscr{Re}\!\left[
\chi\,\eta^{*}\,(1-x)^{1/4}(1-y)^{1/4}(1-z)^{-1/2}
\right] \;=\; 1.
\nn
\end{align}
Here $\chi=|\chi|$ and $\eta=|\eta|e^{i\delta}$ so that $\chi\eta^{*}=|\chi||\eta|e^{-i\delta}$ makes the $\delta$-dependence explicit inside the real parts. All fractional powers are taken on their principal branches (branch cut $(-\infty,0]$) for $(1-x)^{\alpha}$ and $(1-z)^{\alpha}$. We take $r,s\ge 0$ and $\Delta=\theta-\phi$ throughout.

A global minimization at fixed $r=s$ over the relative phase pair $(\Delta,\delta)$ and the amplitude ratio reveals a \emph{universal lower boundary} that the Janus family cannot cross. With optimal phases $(\Delta,\delta)=(\pi,\pi)$ and critical amplitude tuning, the minimum approaches the bound in Eq.~\ref{eq:half_bound} as $r\to 0$, realized asymptotically when the superposition isolates the two-photon sector. Along this optimal boundary the small-$r$ expansion reads
\[
g^{(2)}_{\text{boundary}}(r) \;=\; \tfrac{1}{2} \;+\; \tfrac{3}{4}\,\sinh^{2}\!r \;+\; \mathcal{O}(\sinh^{4}\!r),
\]
while for large squeezing $g^{(2)}_{\text{boundary}}(r)\to 3$, the well-known highly squeezed limit \cite{Loudon2000quantum}. Crucially, there also exists a \emph{practical} local minimum $g^{(2)}\!\approx\!0.567$ at $r\!\approx\!0.34$ (equal squeezing), which defines an operational “sweet spot’’ compatible with state-of-the-art sources and phase stabilization \cite{vahlbruch2016detection,Park2024}.

\begin{figure}[t]
    \centering
    \includegraphics[width=\columnwidth]{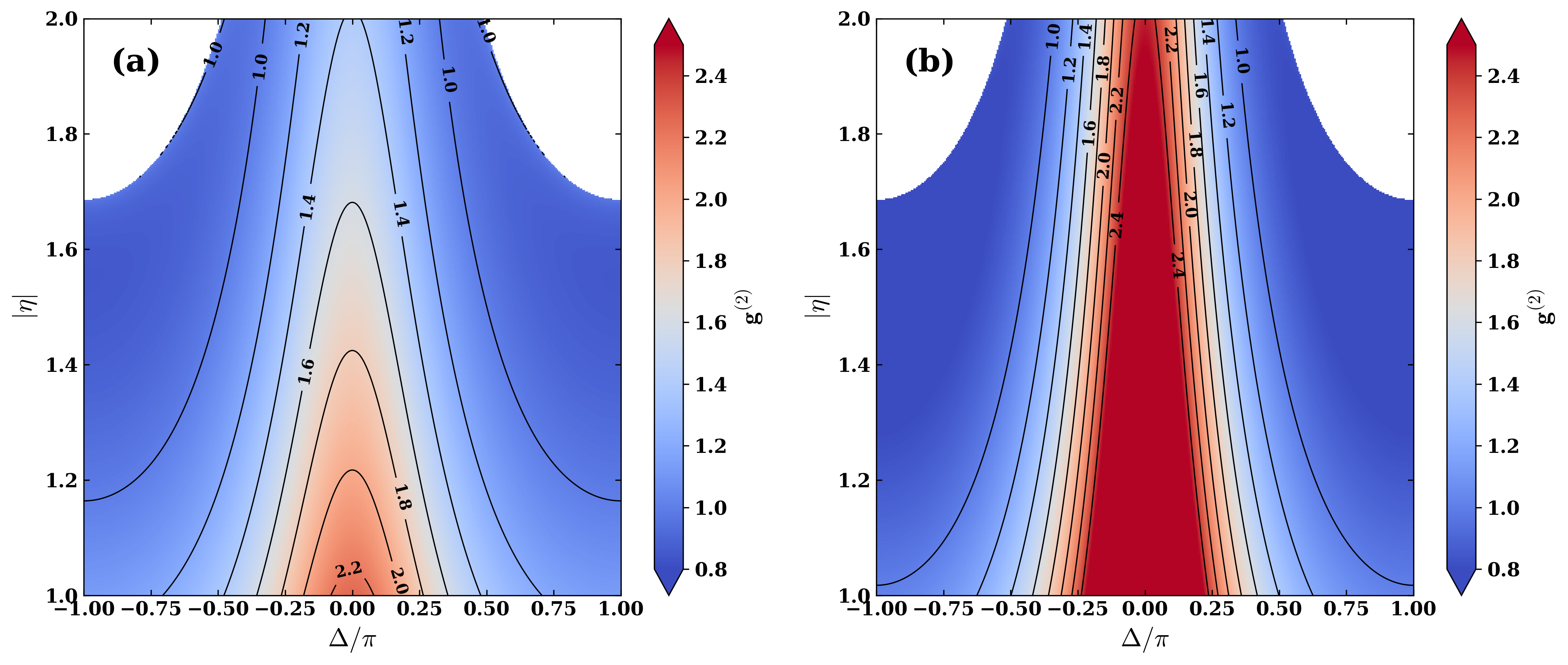}
    \caption{
        \textbf{Unequal squeezing suppresses sub-Poissonian regime.}
        Heatmaps of $g^{(2)}(\psi)$ for fixed $\delta=\pi$ and unequal squeezing: (a) $r=0.7, s=0.3$, (b) $r=0.6, s=0.4$.
        The color scale (right) indicates $g^{(2)}$ values (blue lower, red higher); contours mark selected levels.
        Both cases show predominantly super-Poissonian photon statistics ($g^{(2)}\!>\!1$), with only shallow minima near the classical boundary.
    }
    \label{fig:g2_r_neq_s_scan}
\end{figure}

\begin{figure}[t]
    \centering
    \includegraphics[width=\columnwidth]{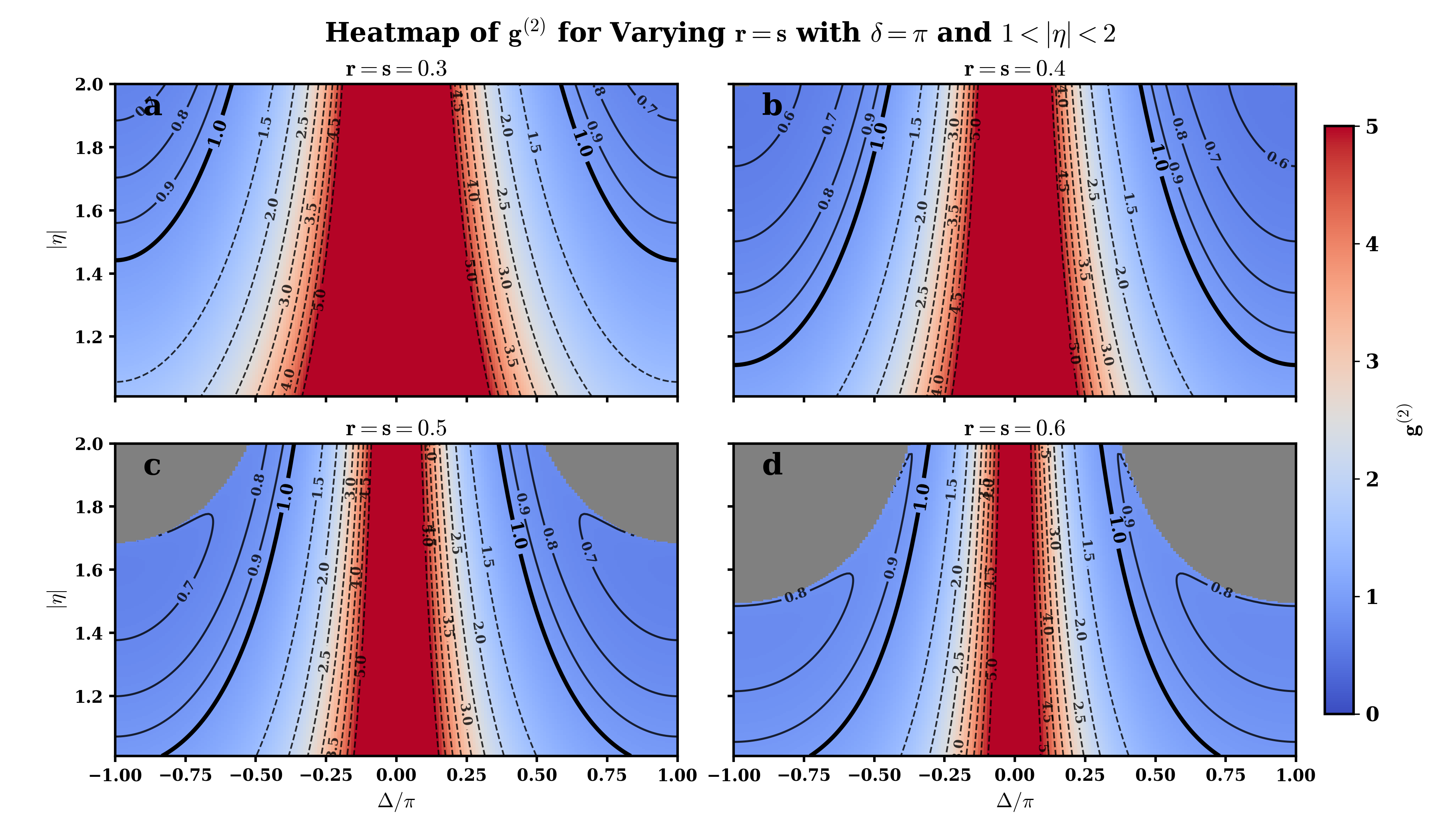}
    \caption{
        \textbf{Phase–amplitude maps for equal squeezing.}
        Heatmaps of $g^{(2)}(\psi)$ for fixed $\delta=\pi$ and $r=s$: (a) $r=0.3$, (b) $r=0.4$, (c) $r=0.5$, (d) $r=0.6$.
        A basin centered near $\Delta\simeq\pi$ emerges and deepens around the practical optimum ($r\!\approx\!0.34$), then gradually shallows at higher squeezing.
    }
    \label{fig:g2_delta_eta_scan}
\end{figure}

The global phase–amplitude landscape clarifies where sub-Poissonian behavior emerges and why symmetry matters. For equal squeezing ($r=s$) and fixed $\delta=\pi$, Fig.~\ref{fig:g2_delta_eta_scan} maps $g^{(2)}(\psi)$ as $r$ increases from $0.3$ to $0.6$. A distinct basin centered near $\Delta\simeq\pi$ develops and deepens up to the practical optimum around $r\!\approx\!0.34$, after which it gradually shallows and contracts. The trough follows a shallow ridge in $|\eta|$ (near $|\eta|\!\approx\!2.2$ at $r\!\approx\!0.34$) where destructive interference suppresses the two-photon amplitude while higher even-photon components persist \cite{Sanders1989Superposition, Buzek1992Superpositions, Obada1997}.

Breaking the squeezing symmetry quickly suppresses the sub-Poissonian regime. Figure~\ref{fig:g2_r_neq_s_scan} shows $(\Delta,|\eta|)$ heat maps for representative unequal pairs $(r,s)=(0.7,0.3)$ and $(0.6,0.4)$ at $\delta=\pi$. In both cases $g^{(2)}(\psi)$ remains predominantly above unity across the scanned domain, with only shallow depressions approaching the classical boundary. This motivates a focus on the symmetric manifold $r=s$ for accessing the sub-Poissonian basin and the local minimum near $r\!\approx\!0.34$.

\begin{figure}[t]
    \centering
    \includegraphics[width=\columnwidth]{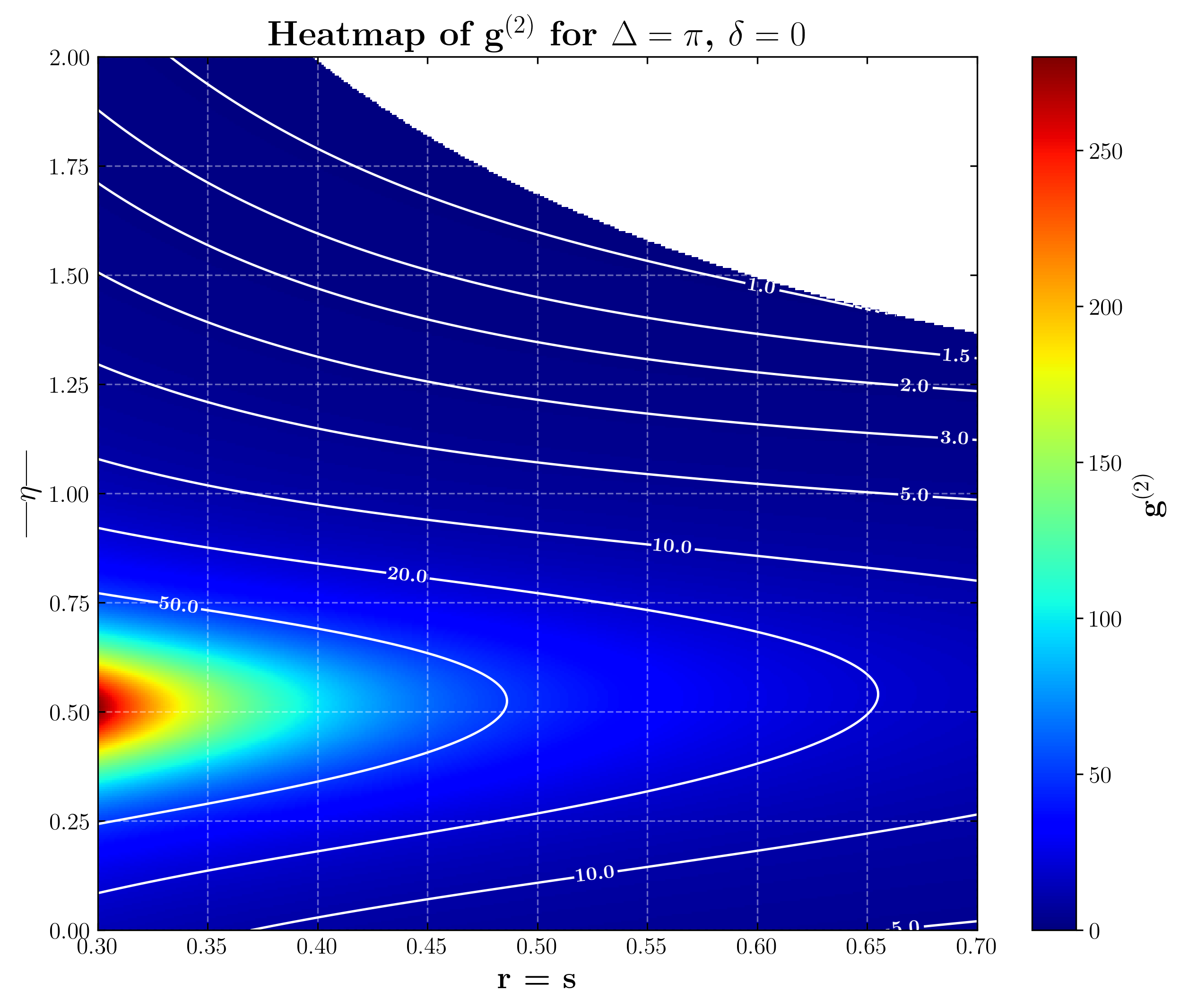}
    \caption{\textbf{Nonoptimal phases:} heat map of $g^{(2)}(\psi)$ for $\Delta=\pi$ and $\delta=0$ versus equal squeezing $r=s$ and superposition amplitude $|\eta|$. Color encodes $g^{(2)}$ (blue lower, red higher); white contours mark selected levels. This configuration is predominantly super-Poissonian ($g^{(2)}>1$).}
    \label{fig:g2_heatmap_delta_pi_delta_0}
\end{figure}

\section{Emergence of sub-Poissonian statistics in symmetric Janus states}

\begin{figure}[t]
    \centering
    \includegraphics[width=\columnwidth]{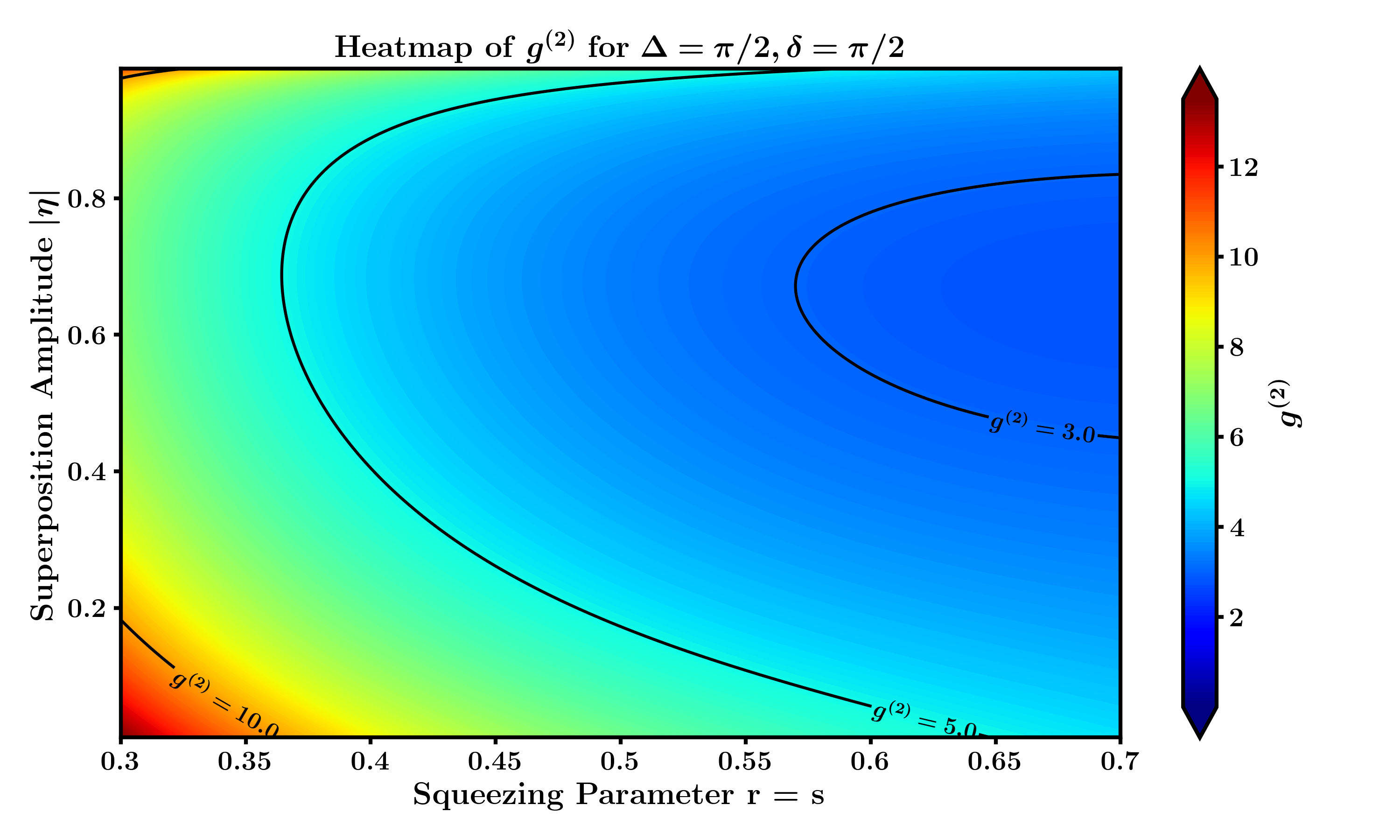}
    \caption{\textbf{Quadrature-offset phases:} heat map of $g^{(2)}(\psi)$ for $\Delta=\pi/2$ and $\delta=\pi/2$ across the same $(r,|\eta|)$ domain. Super-Poissonian photon statistics dominate; no sub-Poissonian statistics are observed within this window.}
    \label{fig:g2_heatmap_delta_pi_2}
\end{figure}

The quantum features of the Janus superposition are most pronounced in the \emph{symmetric} case $r=s$, where the relative phases fully control the photon statistics. Phase scans (Fig.~\ref{fig:g2_delta_eta_scan}) show that $(\Delta,\delta)=(\pi,\pi)$—anti-aligned squeezing orientations with a $\pi$ superposition phase—produce the deepest sub-Poissonian statistics, while generic phases yield predominantly super-Poissonian photon statistics ($g^{(2)}\!>\!1$) \cite{Sanders1989Superposition,Buzek1992Superpositions,Obada1997}; representative nonoptimal cases appear in Figs.~\ref{fig:g2_heatmap_delta_pi_delta_0} and \ref{fig:g2_heatmap_delta_pi_2}.

\begin{figure}[t]
    \centering
    \includegraphics[width=\columnwidth]{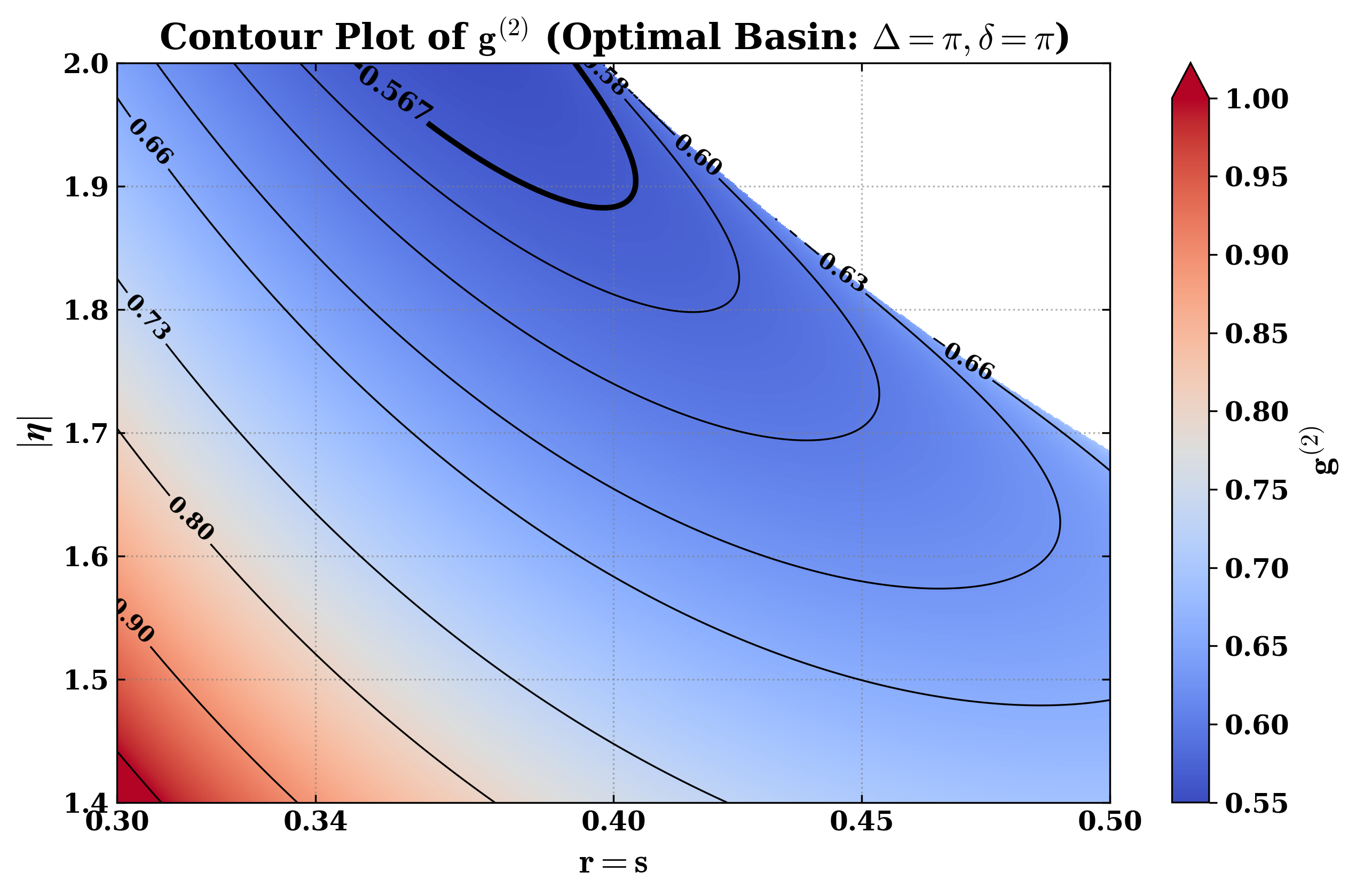}
    \caption{\textbf{sub-Poissonian basin under optimal phases:} contour plot of $g^{(2)}(\psi)$ for $r=s$ as a function of $r$ and $|\eta|$ with $\Delta=\pi$ and $\delta=\pi$. The color scale spans $\approx 0.55$ (deep blue) to $1.0$ (red). A distinct basin with $g^{(2)}\!<\!1$ emerges, reaching a local minimum $\approx 0.567$ near $r\!\approx\!0.34$.}
    \label{fig:g2_heatmap_optimal_antibunching}
\end{figure}

Under the optimal conditions $r=s$, $\Delta=\pi$, and $\delta=\pi$, the general expression for $g^{(2)}(\psi)$ (see SI) simplifies to
\begin{align}
g^{(2)}_{\mathrm{opt}}(\psi)
=& \left( 3 + \frac{1}{\sinh^{2}\! r} \right) \label{eq:g2_optimal_antibunching}\\
& \times
\frac{
1 + 2 K\,|\chi||\eta| + 2 K^{5} |\chi||\eta|\,\big(1 - 2\tanh^{2}\! r \big)
}{
\big( 1 + 2 K\,|\chi||\eta| + 2 K^{3} |\chi||\eta| \big)^{2}
},
\nn
\end{align}
with $K \equiv (1 + 2 \sinh^{2}\! r)^{-1/2}$ and normalization
$|\chi|^{2} + |\eta|^{2} - 2 K\,|\chi||\eta| = 1$.
Equation~\ref{eq:g2_optimal_antibunching} exhibits a broad basin of sub-Poissonian statistics ($g^{(2)}\!<\!1$) (Fig.~\ref{fig:g2_heatmap_optimal_antibunching}), with a practical local minimum $g^{(2)}\!\approx\!0.567$ at $r\!\approx\!0.34$ and $|\eta|\!\approx\!2.20070$, providing an experimentally accessible “sweet spot’’ compatible with state-of-the-art squeezing and phase stabilization \cite{vahlbruch2016detection,Park2024}. Within comparable parameter windows away from $(\pi,\pi)$ no sub-Poissonian statistics are found (Figs.~\ref{fig:g2_heatmap_delta_pi_delta_0},\ref{fig:g2_heatmap_delta_pi_2}); extended ranges and edge cases are discussed in the SI.

\section{Experimental feasibility}

A passive interferometer that mixes two single-mode squeezed vacua implements a linear Gaussian (symplectic) map and therefore sends Gaussian states to Gaussian states; by itself it does not create the coherent single-mode superposition $|\psi\rangle \propto |\xi\rangle + e^{i\delta}|\zeta\rangle$ studied here \cite{weedbrook2012gaussian}. Preparing a Janus superposition in one optical mode requires a minimal non-Gaussian resource—typically a measurement-induced projection (heralding) or an ancilla-assisted weak nonlinearity that coherently selects the squeezing orientation. With that clarification, present-day sources and detectors make an empirical demonstration realistic.

\begin{figure}[t]
    \centering
    \includegraphics[width=\columnwidth]{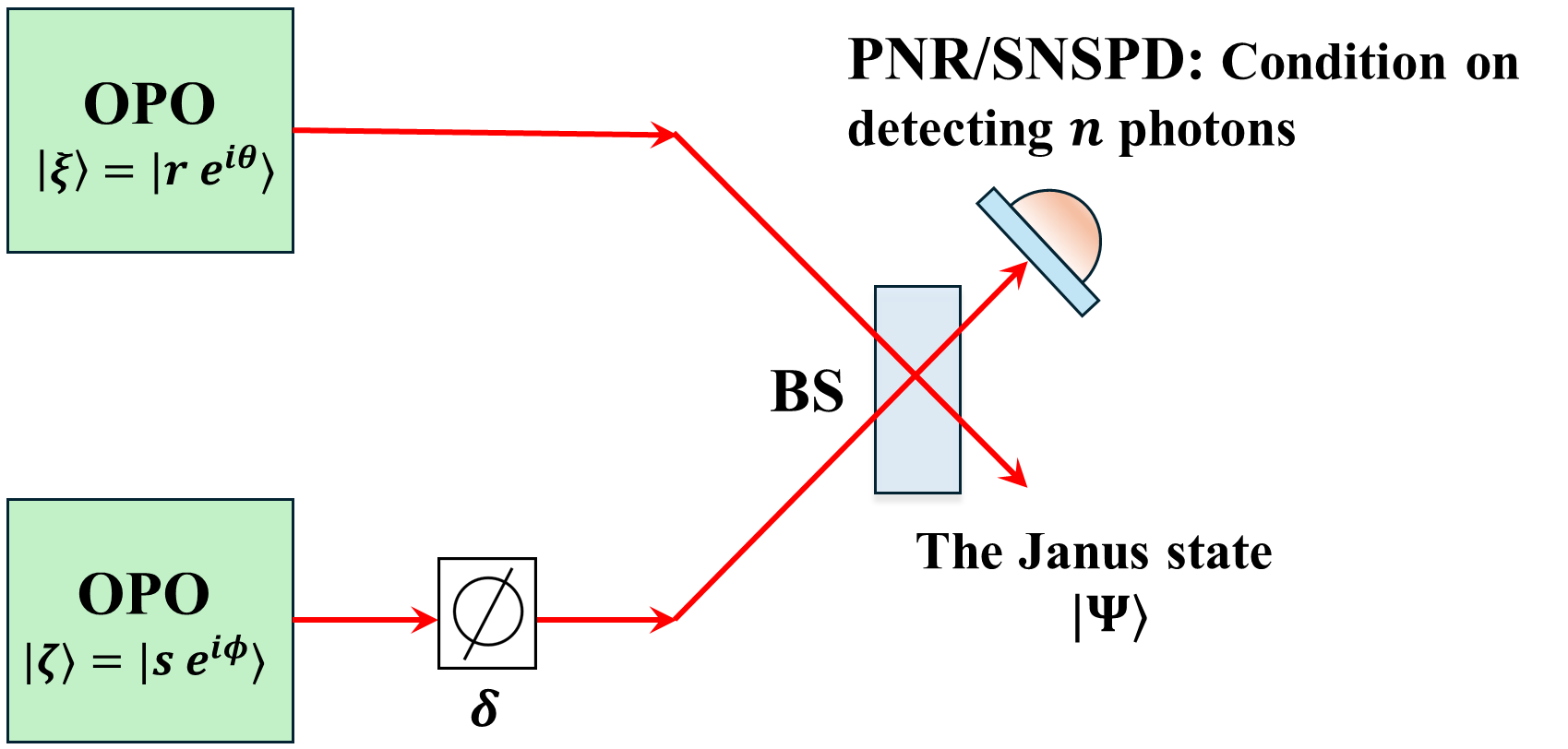}
    \caption{Heralded preparation of a Janus state. Two phase-locked OPOs
    generate single-mode squeezed vacua
    $|\xi\rangle = |r e^{i\theta}\rangle$ and
    $|\zeta\rangle = |s e^{i\phi}\rangle$. The modes interfere on a beam
    splitter (BS). Conditioning on an $n$-photon event in the herald arm
    (PNR/SNSPD) projects the signal arm onto a Janus superposition
    $|\Psi\rangle \propto \chi\,|\xi\rangle + \eta\,e^{i\delta}|\zeta\rangle$.}
    \label{fig:JanusSetup}
\end{figure}

Phase-stable single-mode squeezed vacua at the 10–15\,dB level are routinely generated with below-threshold OPOs, including chip-scale realizations \cite{vahlbruch2016detection,Park2024}, and high-efficiency single-photon and photon-number-resolving detectors enable both faithful $g^{(2)}(0)$ readout and heralding \cite{Zadeh2021,Stasi2023}. A concrete near-term route, sketched in Fig.~\ref{fig:JanusSetup}, mirrors cat-state conditioning \cite{Ourjoumtsev2007,Parigi2007}: two equal-squeezing modes with opposite orientations ($r=s$, $\Delta=\pi$) are combined on a beam splitter (BS), and a click in a high-$\eta$ PNR/SNSPD (or calibrated homodyne) in the herald arm erases which-orientation information and projects the signal arm onto a Janus state $|\psi\rangle \propto \chi|\xi\rangle + \eta e^{i\delta}|\zeta\rangle$, with $\delta$ set by the interferometric phase. Operating near the moderate-squeezing “sweet spot’’ ($r\!\approx\!0.34$) yields practical heralding rates while preserving a broad region of sub-Poissonian photon statistics.

Nonclassical photon statistics are verified with a standard HBT measurement of $g^{(2)}(0)$ \cite{Loudon2000quantum,Zadeh2021,Stasi2023}. Loss, mode mismatch, and phase jitter reduce the depth of the sub-Poissonian dip, but around $r\!\sim\!0.3$–$0.4$ the region with $g^{(2)}(0)<1$ remains robust under realistic imperfections (see SI for tolerance maps and higher-order diagnostics).

\section{Conclusion}
We introduced and analyzed the Janus state—a coherent superposition of two single-mode squeezed vacua—and derived closed-form expressions for its photon-correlation statistics. We established a universal lower bound for the second-order coherence,
$\lim_{r\to 0} g^{(2)}_{\min}(r)=\tfrac{1}{2}$, and identified a practical local minimum $g^{(2)}\!\approx\!0.567$ at $r\!\approx\!0.34$. This elevates sub-Poissonian behavior from an isolated effect to a design rule: interfere Gaussian primitives so that the two-photon amplitude cancels while higher even-photon components persist.

From a metrological perspective, the Janus family provides a tunable single-mode probe for phase estimation. A phase shift generated by the photon-number operator, $U(\varphi)=\exp(-i\varphi a^{\dagger}a)$, has quantum Fisher information $F_{Q}=4(\Delta n)^{2}$, so that $g^{(2)}(0)$ and the Janus parameters jointly control both the mean photon number and its fluctuations, and hence the ultimate phase sensitivity via the quantum Cramér–Rao bound. By varying the relative weights and phases in the Janus superposition one can interpolate between the strongly super-Poissonian squeezed-vacuum regime and a more Fock-like regime with reduced $g^{(2)}(0)$, realizing a trade-off between sub-Poissonian behavior and phase sensitivity. This links our nonclassicality bounds directly to the phase-estimation performance metrics discussed in Refs.~\cite{Sahota2016phase_estimation,Gong2017phase,Chang2022estimation}.

Consistent with experimental reality, linear interferometry alone cannot generate a single-mode superposition of distinct squeezed states; a minimal non-Gaussian step is required. We outlined realistic measurement-induced and ancilla-controlled routes, and showed that the moderate-squeezing “sweet spot’’ offers favorable tolerances to loss and phase noise. Altogether, the Janus construction provides a compact, experimentally accessible blueprint for generating useful nonclassical light from mature Gaussian ingredients, with clear applications to integrated photonic sources, low-light sensing, and hybrid CV/DV information processing.

\begin{backmatter}
\bmsection{Funding} This work was supported by the
Robert A. Welch Foundation (Grant No. A-1261) and the National Science Foundation (Grant No. PHY-2013771).

\bmsection{Acknowledgment} I am grateful to Girish Agarwal, Sergio Cantu, Gershon Kurizki, Mikhail Lukin, Wolfgang Schleich, Marlan Scully, and Suhail Zubairy for discussions.

\bmsection{Disclosures} The authors declare no conflicts of interest.

\bmsection{Data availability} No data were generated or analyzed in the presented research.

\bmsection{Supplemental document}
See Supplement 1 for supporting content.
\end{backmatter}

\onecolumngrid
\appendix
\section*{Supplemental Information for  ``The Janus State: A Universal Lower Bound for Second-Order Coherence''}

\setcounter{equation}{0}
\renewcommand{\theequation}{S\arabic{equation}}

\section{\texorpdfstring{\boldmath What is $g^{(2)}$ of superposition?}{}}

The study of quantum superpositions of non-orthogonal states is a cornerstone of quantum optics, with profound implications for photon statistics and nonclassical light generation. In this section, we address the central question: what is the second-order coherence $g^{(2)}$ for a superposition of two squeezed states? Let us consider two squeezed states, $\ket{\xi} = \ket{r\, e^{i\theta}}$ and $\ket{\zeta} = \ket{s\, e^{i\phi}}$, with arbitrary squeezing parameters and phases. The general form of their superposition can be written as
\begin{align}
|\psi\rangle = \chi \ket{\xi} + \eta \ket{\zeta}.
\label{eq:superposition}
\end{align}
Here, $\chi$ and $\eta$ are complex amplitudes controlling the relative weight and phase of each constituent state.

Before analyzing the superposition, it is instructive to recall the photon statistics of a single squeezed state. For such a state, the well-known result for the normalized second-order correlation function reads
\begin{align}
g^{(2)}(\xi) = \frac{\bra{\xi} a^\dagger a^\dagger a a \ket{\xi}}{\left( \bra{\xi} a^\dagger a \ket{\xi} \right)^2} = 3 + \frac{1}{\sinh^2 r},
\label{eq:g2_single_squeezed}
\end{align}
which demonstrates the enhanced photon bunching ($g^{(2)}>1$) typical of squeezed light. The underlying moments can be made explicit as
\begin{align}
\lVert a^{2}|\xi\rangle \rVert^{2} =& \sinh^2 r \left( 3 \sinh^2 r + 1 \right), 
\qquad
\lVert a \ket{\xi}\rVert^2 = \sinh^2 r
\nn\\
g^{(2)} =& \frac{\lVert a^{2}|\xi\rangle \rVert^{2}}{\lVert a \ket{\xi}\rVert^4} = 3 + \frac{1}{\sinh^2 r}.
\label{eq:single_state_moments}
\end{align}

Turning now to the superposed state $\ket{\psi}$ given by Eq.~(\ref{eq:superposition}), we seek the corresponding second-order coherence $g^{(2)}(\psi)$. Namely,
\begin{align}
g^{(2)}(\psi) = \frac{ \bra{\psi} a^\dagger a^\dagger a a \ket{\psi} }{  \bra{\psi} a^\dagger a \ket{\psi}^2 }.
\label{eq:g2_superposition_def}
\end{align}
This expression involves both diagonal and cross terms in the numerator and denominator, reflecting interference and non-orthogonality between $\ket{\xi}$ and $\ket{\zeta}$.

Let us examine the denominator, which encodes the mean photon number squared for $\ket{\psi}$. It can be conveniently rewritten as
\begin{align}
\bra{\psi} a^\dagger a \ket{\psi}
=\lVert a \ket{\psi}\rVert^2 = \lVert \chi a \ket{\xi}+ \eta  a \ket{\zeta}\rVert^2
= |\chi|^2 \lVert a \ket{\xi}\rVert^2
+ |\eta|^2 \lVert a \ket{\zeta}\rVert^2
+ 2 \mathscr{Re} \left[\chi \eta^* \bra{\zeta} a^\dagger a \ket{\xi}  \right],
\label{eq:mean_photon_number}
\end{align}
where the third term represents interference between the two squeezed states, and can be constructive or destructive depending on the amplitudes and phases.

Next, consider the numerator of $g^{(2)}(\psi)$, which involves the normally-ordered four-operator moment:
\begin{align}
\bra{\psi} a^\dagger a^\dagger a a \ket{\psi}
=&\lVert a^2 \ket{\psi}\rVert^2 = \lVert \chi a^2 \ket{\xi}
+ \eta  a^2 \ket{\zeta}\rVert^2 \nn\\
=& |\chi|^2 \lVert a^{2}|\xi\rangle \rVert^{2}
+ |\eta|^2 \lVert a^2 \ket{\zeta}\rVert^2
+ 2 \mathscr{Re} \left[\chi \eta^* 
\bra{\zeta} a^\dagger a^\dagger a a \ket{\xi}  \right],
\label{eq:second_moment_superposition}
\end{align}
Equation~(\ref{eq:second_moment_superposition}) explicitly captures the interplay between the photon-pair amplitudes in each squeezed state and their mutual coherence.

Taken together, Eqs.~(\ref{eq:g2_superposition_def}), (\ref{eq:mean_photon_number}), and (\ref{eq:second_moment_superposition}) provide a complete and general analytic framework for evaluating $g^{(2)}$ in an arbitrary superposition of two squeezed. As will be discussed in the subsequent sections, the cross terms—arising from non-orthogonality and relative phase—enable quantum statistical behavior that is fundamentally inaccessible to either squeezed state alone. This interference-induced modification of photon correlations is the central physical novelty of this work.

\subsection{\texorpdfstring{\boldmath Cross terms for $g^{(2)}$ of superposition}{}}

A key technical challenge in evaluating the second-order coherence for superpositions of non-orthogonal states is the careful treatment of cross terms. In the context of the superposition state \(\ket{\psi}\) defined in Eq.~(\ref{eq:superposition}), these cross terms arise due to the overlap between the squeezed states \(\ket{\xi}\) and \(\ket{\zeta}\). Specifically, the evaluation of \(g^{(2)}(\psi)\) requires us to compute expectation values such as \(\langle \xi | a^\dagger a | \zeta \rangle\) and \(\langle \xi | a^\dagger a^\dagger a a | \zeta \rangle\), which capture quantum interference between the two constituent states:
\begin{align}
\langle \xi | a^\dagger a | \zeta \rangle \qquad \text{and} \qquad \langle \xi | a^\dagger a^\dagger a a | \zeta \rangle.
\label{eq:cross_term_as}
\end{align}

To proceed, we first recall the standard representations for squeezed states.   These states can be written in the exponential form \cite{Azizi2025Unitary_TFD}:
\begin{align}
|\xi\rangle = (\cosh r)^{-1/2} \exp\left(- \frac{1}{2} \tanh r\, e^{i\theta} a^{\dagger} a^{\dagger} \right) |0\rangle
\nn\\
|\zeta\rangle = (\cosh s)^{-1/2} \exp\left(- \frac{1}{2} \tanh s\, e^{i\phi} a^{\dagger} a^{\dagger} \right) |0\rangle.
\label{eq:squeezed_exp_form}
\end{align}
Here, $r$ and $s$ are the squeezing parameters, while $\theta$ and $\phi$ are the squeezing phases for each state. To simplify notation, we define the parameters \(\alpha = \tanh r\, e^{i\theta}\) and \(\beta = \tanh s\, e^{i\phi}\), encapsulating both the amplitude and phase information. With these definitions, both squeezed states can be expanded explicitly in the Fock basis. This gives:
\begin{align}
| \xi \rangle&=(1-|\alpha|^2)^{1/4} \sum_{n=0}^\infty \frac{1}{n!} \left( -\frac{\alpha}{2} \right)^n \sqrt{(2n)!}\, |2n\rangle \nn\\
| \zeta \rangle&=(1-|\beta|^2)^{1/4} \sum_{m=0}^\infty \frac{1}{m!} \left( -\frac{\beta}{2} \right)^m \sqrt{(2m)!}\, |2m\rangle.
\label{eq:fock_expansion}
\end{align}
Equation~(\ref{eq:fock_expansion}) shows that squeezed states consist only of even photon number components, with coefficients determined by the parameters $\alpha$ and $\beta$.

\paragraph{Normalization and Overlap.}  
The calculation of the cross terms, and in particular the normalization of $\ket{\psi}$, crucially depends on the overlap \(\langle \zeta | \xi \rangle\) between the two squeezed states. This overlap, given their non-orthogonality, does not vanish and can be computed by direct substitution from Eq.~(\ref{eq:fock_expansion}):
\begin{align}
\langle \zeta  | \xi \rangle =& (1-x)^{1/4}(1-y)^{1/4}
\sum_{m=0}^\infty \frac{1}{m!} \left( -\frac{\beta^*}{2} \right)^m \sqrt{(2m)!}\, 
\sum_{n=0}^\infty \frac{1}{n!} \left( -\frac{\alpha}{2} \right)^n \sqrt{(2n)!}\, \langle 2m |2n\rangle \nn\\
=& (1-x)^{1/4}(1-y)^{1/4}
\sum_{n=0}^\infty \frac{(2n)!}{(n!)^2} \left( \frac{\alpha\beta^*}{4} \right)^n \nn\\
=& (1-x)^{1/4}(1-y)^{1/4}
\sum_{n=0}^\infty \frac{(2n-1)!!}{(2n)!!} \left( \alpha\beta^*\right)^n \nn\\
=& (1-|\alpha|^2)^{1/4}(1-|\beta|^2)^{1/4} (1-\alpha\beta^*)^{-1/2},
\label{eq:overlap_series}
\end{align}
where,
\begin{align}
x =& |\alpha|^2 = \tanh^2 r, \qquad (1-x)^{1/4} = (\cosh r)^{-1/2} \nn\\
y =& |\beta|^2 = \tanh^2 s, \qquad (1-y)^{1/4} = (\cosh s)^{-1/2} \nn\\
z =& \alpha \beta^* = \tanh r\, \tanh s\, e^{i(\theta-\phi)},
\label{eq:x_y_z_def}
\end{align}
and we have made use of the combinatorial identity
\begin{equation}
(1-z)^{-1/2} = \sum_{n=0}^{\infty} \frac{(2n-1)!!}{(2n)!!} z^n.
\label{eq:comb_identity}
\end{equation}
Equation~(\ref{eq:overlap_series}) provides a closed-form expression for the overlap in terms of the squeezing parameters and phases. Moreover, note that if $\xi=\zeta$, then $\alpha=\beta$, and hence Equation~(\ref{eq:overlap_series}) gives $\langle \xi  | \xi \rangle=1$

Consequently, the normalization of the superposed state $\ket{\psi}$, which must satisfy $\langle \psi|\psi\rangle = 1$, becomes
\begin{align}
\langle \psi|\psi\rangle =& |\chi|^2 \langle \xi|\xi\rangle 
+ |\eta|^2  \langle \zeta|\zeta\rangle
+2 \mathscr{Re}\Big[\chi\eta^* \langle \zeta  | \xi \rangle \Big] 
 \nn\\
=&  |\chi|^2 + |\eta|^2  
+2 \mathscr{Re}\Big[\chi\eta^* 
 (1-x)^{1/4}(1-y)^{1/4} (1-z)^{-1/2}
 \Big],
\label{eq:psi_norm_full}
\end{align}
since, by construction, each squeezed state is normalized individually. Summarizing, the final normalization constraint takes the compact form
\begin{align}
\boxed{\quad|\chi|^2 + |\eta|^2  
+2 \mathscr{Re}\Big[\chi\eta^* 
 (1-x)^{1/4}(1-y)^{1/4} (1-z)^{-1/2}
 \Big]=1\,.
 \quad}
\label{eq:psi_norm_gen}
\end{align}

\subsection{\texorpdfstring{\boldmath Finding $\langle \zeta | a^\dagger  a | \xi \rangle$}{}}

A central step in the evaluation of the second-order coherence function for a superposition of squeezed states is the determination of off-diagonal expectation values, in particular $\langle \zeta | a^\dagger a | \xi \rangle$. This term encodes the quantum interference between the two constituent squeezed states in the observable photon statistics.

We begin with the explicit expression for the matrix element in operator form:
\begin{align}
\langle \zeta| a^\dagger a |\xi  \rangle = (\cosh r)^{-1/2} (\cosh s)^{-1/2} \langle 0 |
\exp\left( -\frac{1}{2} \beta^* a a \right)  a^\dagger a
\exp\left( -\frac{1}{2} \alpha a^{\dagger}  a^{\dagger} \right)
|0\rangle,
\label{eq:adag_a_cross}
\end{align}
where we have written both squeezed in their exponential disentangled forms, following Eq.~(\ref{eq:squeezed_exp_form}).

To simplify the calculation, we first act with the annihilation operator $a$ on the squeezed $\ket{\xi}$. As shown by direct differentiation, this yields:
\begin{align}
a \exp\left( -\frac{1}{2} \alpha a^{\dagger} a^{\dagger} \right) |0\rangle = \frac{\partial}{\partial a^{\dagger}} \exp\left( -\frac{1}{2} \alpha a^{\dagger} a^{\dagger} \right) |0\rangle = -\alpha a^{\dagger} \exp\left( -\frac{1}{2} \alpha a^{\dagger} a^{\dagger} \right) |0\rangle.
\label{eq:a_on_squeeze}
\end{align}
Analogously, acting with $a^\dagger$ to the left on $\langle 0| \exp(-\frac{1}{2} \beta^* a a)$ results in:
\begin{align}
\langle 0 | \exp\left( -\frac{1}{2} \beta^* a a \right) a^\dagger = -\beta^* \langle 0 | \exp\left( -\frac{1}{2} \beta^* a a \right) a.
\label{eq:adag_on_left}
\end{align}
Combining these results, Eq.~(\ref{eq:adag_a_cross}) reduces to
\begin{align}
\langle \zeta| a^\dagger a |\xi  \rangle = (\cosh r)^{-1/2} (\cosh s)^{-1/2} \alpha \beta^* \langle 0 | \exp(-\frac{1}{2} \beta^* a a) a a \exp(-\frac{1}{2} \alpha a^{\dagger} a^{\dagger}) | 0 \rangle.
\label{eq:reduced_cross}
\end{align}
At this stage, it is advantageous to adopt a Fock basis expansion for both squeezed states, as in Eq.~(\ref{eq:fock_expansion}), and then evaluate the matrix elements by summing over photon number. To this end, we note that:
\begin{align}
a \ket{\xi} 
&= (1-x)^{1/4} \left( -\alpha \right) a^{\dagger} \exp\left( -\frac{\alpha}{2} a^{\dagger} a^{\dagger} \right) |0\rangle \nn\\
&= -\alpha (1-x)^{1/4} \sum_{n=0}^{\infty} \frac{1}{n!} \left(-\frac{\alpha}{2}\right)^n \sqrt{(2n+1)!}\, |2n+1\rangle,
\label{eq:a_xi_fock}
\end{align}
and, for the bra state,
\begin{align}
\langle \zeta | a^\dagger 
= (1-y)^{1/4} \left( -\beta^* \right) \sum_{m=0}^{\infty} \frac{1}{m!} \left( -\frac{\beta^*}{2} \right)^m \sqrt{(2m+1)!} \langle 2m+1 |,
\label{eq:adag_zeta_fock}
\end{align}

The matrix element can now be explicitly evaluated by the orthonormality of the Fock basis:
\begin{align}
\langle \zeta | a^\dagger a | \xi \rangle
=& \alpha \beta^* (1-x)^{1/4} (1-y)^{1/4} \nn\\
&\times
\sum_{m,n=0}^{\infty} \frac{1}{m! n!}
\left(- \frac{\alpha}{2} \right)^n \left(- \frac{\beta^*}{2} \right)^m \sqrt{(2n+1)!} \sqrt{(2m+1)!} \langle 2m+1 | 2n+1 \rangle \nn\\
=& \alpha \beta^* (1-x)^{1/4} (1-y)^{1/4} \sum_{n=0}^{\infty} \frac{1}{(n!)^2} \left( \frac{\alpha \beta^*}{4} \right)^n (2n+1)!,
\label{eq:adag_a_xi_zeta_sum}
\end{align}
where the last equality follows since $\langle 2m+1 | 2n+1 \rangle = \delta_{mn}$, thus collapsing the double sum.

Next, recognizing the structure of the series, we introduce the variable $z = \alpha \beta^*$, and use the known result for the generating function of factorial moments. Thus, the summation simplifies to:
\begin{align}
\langle \zeta | a^\dagger a | \xi \rangle
= \frac{ (1-x)^{1/4} (1-y)^{1/4} }{ (1-z)^{1/2} }
\, 
\underbrace{
z (1-z)^{1/2} 
\sum_{n=0}^\infty (2n+1) \frac{(2n-1)!!}{(2n)!!} z^n
}_{\displaystyle \frac{z}{1-z}},
\label{eq:cross_term_a_gen_func}
\end{align}
where we have used the closed-form expression for the sum, which will be derived in Eq.~(\ref{eq:closed_form_n}), yielding $\displaystyle\frac{z}{1-z}$. Collecting all prefactors, we arrive at the following result:
\begin{align}
\boxed{ \quad
\langle \zeta | a^\dagger a | \xi \rangle
= \frac{ (1-x)^{1/4} (1-y)^{1/4} }{ (1-z)^{3/2} }\, z\,. \quad
}
\label{eq:cross_term_a}
\end{align}
\vspace{.1cm}
\subsubsection{Finding closed forms for series}
Using Taylor series, one can show that
\begin{align}
(1-z)^{-1/2} = \sum C_n z^n, \qquad
C_n = \frac{(2n-1)!!}{(2n)!!}.
\end{align}
Then we have
\begin{align}
z \frac{\partial}{\partial z} (1-z)^{-1/2} = \sum C_n n z^n
, \qquad
z \frac{\partial}{\partial z} z \frac{\partial}{\partial z} (1-z)^{-1/2} = \sum C_n n^2 z^n. \label{eq:series}
\end{align}
Moreover, 
\begin{align}
z \frac{\partial}{\partial z} (1-z)^{-1/2} =& \frac{1}{2} z (1-z)^{-3/2}, \nn\\
z \frac{\partial}{\partial z}
\Big( z \frac{\partial}{\partial z} (1-z)^{-1/2}\Big)
=& \frac{1}{2} z \frac{\partial}{\partial z} \Big(z (1-z)^{-3/2}\Big)
= \frac{1}{2} z \left( (1-z)^{-3/2} + \frac{3}{2} z (1-z)^{-5/2} \right). \label{eq:derivatives}
\end{align}
 Now we have all ingredients to find the closed form fo the expressions appear in our calculation.

\paragraph{\boldmath Closed form for $\sum_{n=0}^\infty (2n+1)\bm{C}_n z^n$.}
\vspace{.2cm}
Thus, using Eqs.~(\ref{eq:series},\ref{eq:derivatives})  we have
\begin{align}
z(1-z)^{1/2} \sum_{n=0}^\infty z^n (2n+1) C_n =&
z(1-z)^{1/2}\big( 2z \frac{\partial}{\partial z} +1 \big) (1-z)^{-1/2} \nn\\
=&
z(1-z)^{1/2}\big(  \frac{z}{1-z} +1 \big) (1-z)^{-1/2}
= \frac{z}{1-z}. \label{eq:closed_form_n}
\end{align}

\vspace{.2cm}
\paragraph{\boldmath Closed form for $\sum_{n=0}^\infty (2n+1)^2\bm{C}_n z^n$.}
Again, using Eqs.~(\ref{eq:series}, \ref{eq:derivatives})  we have
\begin{align}
 z(1-z)^{1/2}& \sum_{n=0}^\infty (2n+1)^2 C_n z^n \nn\\
 &= z(1-z)^{1/2} \left( 4z \frac{\partial}{\partial z} z \frac{\partial}{\partial z} + 4z \frac{\partial}{\partial z} + 1 \right) (1-z)^{-1/2} \nn\\
 &= z(1-z)^{1/2} \Bigg\{ 2 z \left( (1-z)^{-3/2} + \frac{3}{2} z (1-z)^{-5/2} \right) 
+ 2 z (1-z)^{-3/2}+ (1-z)^{-1/2} \Bigg\}\nn\\
&= \frac{z(2z+1)}{(1-z)^2}. \label{eq:closed_form_n^2}
\end{align}

\vspace{.2cm}

\subsection{\texorpdfstring{\boldmath Finding $\langle \zeta | a^\dagger a^\dagger a a | \xi \rangle$}{}}

Having determined the cross term involving a single pair of creation and annihilation operators, we now proceed to compute the more intricate expectation value $\langle \zeta | a^\dagger a^\dagger a a | \xi \rangle$. 

The approach is analogous to the previous subsection, but with an additional application of ladder operators. We begin by acting twice with the annihilation operator $a$ on the squeezed $|\xi\rangle$. Utilizing the Fock state expansion Eq.~(\ref{eq:a_xi_fock}), $a^2 \ket{\xi}$ reads
\begin{align}
a^2 \ket{\xi}
=& a \left( a \ket{\xi} \right)
= -\alpha (1-x)^{1/4} a \left( \sum_{n=0}^\infty \frac{1}{n!} \left( -\frac{\alpha}{2} \right)^n \sqrt{(2n+1)!}\, |2n+1\rangle \right) \nn\\
=& -\alpha (1-x)^{1/4} \sum_{n=0}^\infty \frac{1}{n!} \left( -\frac{\alpha}{2} \right)^n \sqrt{(2n+1)!} \sqrt{2n+1} |2n\rangle,
\label{eq:annihilate_twice}
\end{align}
where, we have simply used $a |2n+1\rangle = \sqrt{2n+1} |2n\rangle $. For the bra state, we similarly apply $a^\dagger a^\dagger$ to the squeezed $\langle \zeta|$, yielding:
\begin{align}
\langle \zeta | a^\dagger a^\dagger
= -\beta^* (1-y)^{1/4} \sum_{m=0}^\infty \frac{1}{m!} \left( -\frac{\beta^*}{2} \right)^m \sqrt{(2m+1)!} \sqrt{2m+1} \langle 2m |.
\label{eq:bra_creation_twice}
\end{align}
We are now in a position to assemble the full cross term. Taking the inner product, and invoking the orthonormality of the Fock basis, we find:
\begin{align}
\langle \zeta | (a^\dagger)^2 a^2 | \xi \rangle
=& \alpha \beta^* (1-x)^{1/4} (1-y)^{1/4} \nn\\
&\times \sum_{n,m=0}^\infty \frac{1}{n! m!}
\left( -\frac{\alpha}{2} \right)^n
\left( -\frac{\beta^*}{2} \right)^m
\sqrt{(2n+1)! (2m+1)!} \sqrt{2n+1}\sqrt{2m+1}
\langle 2m | 2n \rangle \nn\\
=& \alpha \beta^* (1-x)^{1/4} (1-y)^{1/4} \sum_{n=0}^\infty \frac{1}{(n!)^2} \left( \frac{\alpha \beta^*}{4} \right)^n (2n+1)! (2n+1),
\label{eq:cross_fourth_order_sum}
\end{align}
where, as before, $\langle 2m | 2n \rangle = \delta_{mn}$ reduces the sum to a single index. Since 
\begin{align}
\frac{1}{4^n (n!)^2}(2n+1)! (2n+1)=&
(2n+1)^2 \frac{(2n)!!(2n-1)!!}{2^{2n}(n!)^2}=
(2n+1)^2 \frac{2^n n!(2n-1)!!}{2^{2n}(n!)^2} \nn\\
=&(2n+1)^2 \frac{(2n-1)!!}{(2n)!!},
\end{align}
using the closed form found in Eq.~(\ref{eq:closed_form_n^2}) for $z=\alpha \beta^*$, we have
\begin{align}
z (1-z)^{1/2} \sum_{n=0}^\infty (2n+1)^2 \frac{(2n-1)!!}{(2n)!!} z^n = \frac{z (2z+1)}{(1-z)^2}.
\end{align}
Putting all factors together, the final closed-form expression is obtained:
\begin{align}
\boxed{\quad
\langle \zeta | a^\dagger a^\dagger a a | \xi \rangle
= \frac{(1-x)^{1/4} (1-y)^{1/4} z (2z+1)}{(1-z)^{5/2}}.
\quad}
\label{eq:fourth_order_cross}
\end{align}
where $x$, $y$, and $z$ are defined in Eq.~(\ref{eq:x_y_z_def}).

\subsection{\texorpdfstring{\boldmath Finding $g^{(2)}$ of squeezed state}{}}
Finally, let us determine the well-known second-order coherence, $g^{(2)}$, for a single squeezed state. This is achieved by setting $\xi=\zeta$ in Eq~(\ref{eq:cross_term_a}). Consequently, Eq~(\ref{eq:x_y_z_def}) yields $x=y=z=\abs{\alpha}^2$, and we find
\begin{align}
\langle \xi | a^\dagger a | \xi \rangle
= \frac{x}{1-x}.
\label{eq:squeezed_expc_a}
\end{align}
Moreover, Eq~(\ref{eq:fourth_order_cross}) reduces to the following simple form for $\xi=\zeta$ (i.e., $x=y=z$):
\begin{align}
\langle \xi | a^\dagger a^\dagger a a | \xi \rangle
= \frac{x(2x+1)}{(1-x)^2}.
\label{eq:squeezed_expc_a^2}
\end{align}
Hence, $g^{(2)}$ for a single squeezed state is then
\begin{align}
g^{(2)}(\xi)= \frac{\langle \xi | a^\dagger a^\dagger a a | \xi \rangle}
{\langle \xi | a^\dagger a | \xi \rangle^2}=
\frac{2x+1}{x}=2+\coth^2{r}=3+\frac{1}{\sinh^2{r}}.
\end{align}

\subsection{Plugging into the superposition formulas}
To systematically analyze the photon statistics of a superposition of two squeezed states, it is essential to first compile the key expectation values for a single squeezed state. These quantities provide the building blocks for the more general calculations.

For a squeezed state $|\xi\rangle$ (with squeezing parameter $r$ and phase $\theta$), we have:

\begin{align}
\lVert a\ket{\xi}\rVert^2 =& \sinh^2 r = \frac{x}{1-x},
\qquad\qquad
\lVert a^2\ket{\xi}\rVert^2 = \frac{x(2x+1)}{(1-x)^2},
\nn\\
\lVert a\ket{\zeta}\rVert^2  =& \sinh^2 s = \frac{y}{1-y},
\qquad\qquad
\lVert a^2\ket{\zeta}\rVert^2 = \frac{y(2y+1)}{(1-y)^2}, \nn\\
\langle \zeta | a^\dagger a | \xi \rangle =& \frac{(1-x)^{1/4}(1-y)^{1/4} z}{(1-z)^{3/2}}, \nn\\
\langle \zeta | a^\dagger a^\dagger a a | \xi \rangle =& \frac{(1-x)^{1/4}(1-y)^{1/4} z(2z+1)}{(1-z)^{5/2}},
\label{eq:sq_summary}
\end{align}
where $x, y, z$ are given by Eq.~(\ref{eq:x_y_z_def}). These expressions capture both the diagonal and off-diagonal (cross) terms necessary for photon correlation functions.

These results allow us to efficiently plug the relevant quantities into the superposition formulas. In particular, the normalization and photon number moments for the superposed state $|\psi\rangle = \chi|\xi\rangle + \eta|\zeta\rangle$ can be written in closed analytic form. The mean photon number and its higher moments for $|\psi\rangle$ take the forms:
\begin{align}
\lVert a \ket{\psi}\rVert^2
=& |\chi|^2 \frac{x}{1-x}
+ |\eta|^2 \frac{y}{1-y}
+ 2\, \mathscr{Re} \left[ \chi \eta^* \frac{(1-x)^{1/4}(1-y)^{1/4} z}{(1-z)^{3/2}} \right]
\label{eq:superposition__moment_a}\\
\lVert a^2 \ket{\psi}\rVert^2
=& |\chi|^2 \frac{x(2x+1)}{(1-x)^2}
+ |\eta|^2 \frac{y(2y+1)}{(1-y)^2}
+ 2\, \mathscr{Re} \left[ \chi \eta^* \frac{(1-x)^{1/4}(1-y)^{1/4} z(2z+1)}{(1-z)^{5/2}} \right]
\label{eq:superposition__moment_a^2}
\end{align}
These formulas generalize the photon statistics for superposed squeezed states, capturing quantum interference between the two components. Finally $g^{(2)}(\psi)$ can be found as:
\begin{align}
\boxed{\quad g^{(2)} = \frac{ |\chi|^2 \displaystyle\frac{x(2x+1)}{(1-x)^2}
+ |\eta|^2 \frac{y(2y+1)}{(1-y)^2}
+ 2\, \mathscr{Re} \left[ \chi \eta^* \frac{(1-x)^{1/4}(1-y)^{1/4} z(2z+1)}{(1-z)^{5/2}} \right] }
{\Bigg\{|\chi|^2 \displaystyle \frac{x}{1-x}
+ |\eta|^2 \frac{y}{1-y}
+ 2\, \mathscr{Re} \left[ \chi \eta^* \frac{(1-x)^{1/4}(1-y)^{1/4} z}{(1-z)^{3/2}} \right] \Bigg\}^2 }, \quad}
\label{eq:g2_final_def}
\end{align}
\vspace{.1cm}
with the normalization given in Eq~(\ref{eq:psi_norm_gen}).
\section{Simplification: equal squeezing, phase offset}
An important special case arises when both squeezed states have the same squeezing strength but differ in phase: $r = s$, $\theta \neq \phi$. This regime is particularly relevant for experiments in which the phase between two identical squeezed modes is controlled. The simplifications are:
\begin{align}
x = y = \tanh^2 r,
\quad 
(1-x)^{1/4}= (1-y)^{1/4} = (\cosh r)^{-1/2},
\quad 
z = x e^{i\Delta},
\label{eq:same_squeezing_param}
\end{align}
where $\Delta = \theta-\phi$ is the relative phase. 

Now, the mean photon number in the superposed state Eq~(\ref{eq:superposition__moment_a}) reduces to
\begin{align}
\lVert a \ket{\psi}\rVert^2
=&  \frac{x}{1-x} \Big(|\chi|^2+ |\eta|^2 \Big)
+ 2\, (1-x)^{1/2}\mathscr{Re} \left[ \chi \eta^* \frac{z}{(1-z)^{3/2}} \right] \nn\\
=& \frac{x}{1-x} \Bigg\{ |\chi|^2+ |\eta|^2
+ 2\, (1-x)^{3/2}\mathscr{Re} \left[  \frac{\chi \eta^* e^{i\Delta}}{(1-z)^{3/2}} \right]
\Bigg\}
\label{eq:a_psi_simplified}
\end{align}
where the curly bracket groups the direct terms and the interference/cross terms, which encode the effect of quantum superposition. The second moment, the expression Eq~(\ref{eq:superposition__moment_a^2}) becomes:
\begin{align} 
\lVert a^2 \ket{\psi}\rVert^2
=&  \frac{x(2x+1)}{(1-x)^2} \Big(|\chi|^2+ |\eta|^2 \Big)
+ 2\, x (1-x)^{1/2}\mathscr{Re} \left[ \chi \eta^* \frac{e^{i\Delta}(2z+1)}{(1-z)^{5/2}} \right] \nn\\
=& \frac{x(2x+1)}{(1-x)^2} \Bigg\{ |\chi|^2+ |\eta|^2
+ 2\, (1-x)^{5/2}\mathscr{Re} 
\left[  \frac{\chi \eta^* e^{i\Delta} }{(1-z)^{5/2}} \displaystyle \frac{2z+1}{2x+1} \right]
\Bigg\}.
\label{eq:a2_psi_simplified}
\end{align}
The normalization of the superposed state Eq~(\ref{eq:psi_norm_gen}) also takes a compact form:
\begin{align}
\boxed{\quad|\chi|^2 + |\eta|^2  
+2 \mathscr{Re}\Big[\chi\eta^* 
 \Big(\frac{1-x}{1-x e^{i\Delta}} \Big)^{1/2}
  \Big]=1\,.
 \quad}
\label{eq:psi_normalization_r=s}
\end{align}
This ensures that $|\psi\rangle$ is a normalized quantum state for any choice of superposition coefficients. Consequently, the second-order coherence function for the superposed state simplifies to:
\begin{align}
\boxed{\quad 
g^{(2)}(\psi) 
= \frac{2x+1}{x}\,
\frac{\Bigg\{ |\chi|^2+ |\eta|^2
+ 2\, \mathscr{Re} 
\left[\chi \eta^*  e^{i\Delta} \Big(\displaystyle \frac{2z+1}{2x+1}\Big) \,
\displaystyle \Big(\frac{1-x}{1-z}\Big)^{5/2} \right]
\Bigg\}}
{\Bigg\{ |\chi|^2+ |\eta|^2
+ 2\, \mathscr{Re} \left[\chi \eta^* e^{i\Delta}\,
\displaystyle \Big(\frac{1-x}{1-z}\Big)^{3/2} \right]
\Bigg\}^2}\,.
\quad }
\label{eq:g2_superposed_simplified_SM}
\end{align}

\vspace{.3cm}
\paragraph{Elementary algebra.} Let's perform some elementary algebraic manipulations in the following equations:
\begin{align}
\frac{x(2x+1)}{(1-x)^2} = \frac{\tanh^2 r \left(2\tanh^2 r + 1\right)}{(1-\tanh^2 r)^2}
= \cosh^4 r \frac{\sinh^2 r}{\cosh^2 r} \left(\frac{2\sinh^2 r + \cosh^2 r}{\cosh^2 r}\right)
\end{align}
Hence
\[
\frac{x(2x+1)}{(1-x)^2} = \sinh^2 r\, (3\sinh^2 r + 1).
\]

Furthermore, $\displaystyle\frac{1-x}{1-z}$, which appears in nearly all cross-terms between the two squeezed states in the superposition, can be simplified as follows.  Since $x = \tanh^2 r$ and $z = x e^{i\Delta}$ (for phase difference $\Delta$), we can write
\begin{align}
\frac{1-x}{1-z}=&\frac{1-x}{1-x\, e^{i\Delta}}=\frac{1}{\cosh^2 r- \sinh^2 r\, e^{i\Delta} }
\end{align}
The denominator can be written as:
\begin{align}
\cosh^2 r- \sinh^2 r\, e^{i\Delta}=&  \nn\\
=& (\cosh^2 r - \sinh^2 r \cos\Delta) - i\sinh^2 r \sin\Delta \nn\\
=& \Big( (\cosh^2 r - \sinh^2 r \cos\Delta)^2
+ (\sinh^2 r \sin\Delta)^2 \Big)^{1/2} \nn\\
&\times
\exp{-i \tan^{-1} \left( \frac{\sinh^2 r \sin\Delta}
{\cosh^2 r - \sinh^2 r \cos\Delta} \right) }
\end{align}
Thus, the denominator can be expressed as a product of a modulus (a real positive factor) and a pure phase, which we label $f(r,\Delta)^{1/2}$ and $e^{i\gamma}$, respectively:
\begin{align}
\cosh^2 r- \sinh^2 r\, e^{i\Delta}
=& f(r,\Delta)^{1/2}
e^{i\gamma}.
\end{align}
This leads to the identity,
\begin{align}
\boxed{ \quad \frac{1-x}{1-z}=f(r,\Delta)^{-1/2}
e^{-i\gamma},
\quad}
\label{eq:fxgamma}
\end{align}
where
\begin{align}
\boxed{\quad f(r,\Delta) = \cosh^4 r - 2\sinh^2 r \cosh^2 r \cos\Delta
+ \sinh^4 r, \quad 
\gamma=- \tan^{-1} \left( \frac{ \sin\Delta}
{\coth^2 r -\cos\Delta} \right).  \quad}
\label{eq:fgamma}
\end{align}
\paragraph{Implications for normalization and $g^{(2)}(\psi)$:}
With these simplifications, every normalization and photon correlation term involving $\displaystyle\frac{1-x}{1-z}$ can now be systematically recast in terms of $f(r,\Delta)$ and $\gamma$, leading to substantial analytical and numerical advantages. The normalization condition Eq~(\ref{eq:psi_normalization_r=s}) becomes
\begin{align}
|\chi|^2 + |\eta|^2  
+2 f(r,\Delta)^{-1/4} \mathscr{Re}\Big[\chi\eta^*  
e^{-\frac{i \gamma}{2}}
  \Big]=1\,. \label{eq:psi_normalization_r=s_f}
 \end{align}
Substituting these analytic expressions into the original formula for $g^{(2)}(\psi)$, we arrive at the following general and compact result:
\begin{align}
g^{(2)}(\psi) 
=& \frac{2x+1}{x}\, \Bigg\{ 1-2 f(r,\Delta)^{-1/4} \mathscr{Re}\Big[\chi\eta^*  
e^{-\frac{i \gamma}{2}}
  \Big]
+ 2\,f(r,\Delta)^{-3/4} \mathscr{Re} \left[\chi \eta^*
e^{i\big(\Delta-\frac32\gamma\big)}\,
\right]
\Bigg\}^{-2} \nn\\
&
\Bigg\{ 1-2 f(r,\Delta)^{-1/4} \mathscr{Re}\Big[\chi\eta^*  
e^{-\frac{i \gamma}{2}}
  \Big]
+ 2\, f(r,\Delta)^{-5/4} \mathscr{Re} 
\left[\chi \eta^*  e^{i\big(\Delta-\frac52 \gamma\big)}
\Big(\displaystyle \frac{2x e^{i\Delta}+1}{2x+1}\Big)
 \right]
\Bigg\}
\,.
\end{align}

\subsection{\texorpdfstring{\boldmath Finding $g^{(2)}$ in terms of the phase of coefficients}{}}

To further simplify and parametrize the superposition state, it is advantageous to express the coefficients $\chi$ and $\eta$ in terms of their magnitudes and relative phase. Without loss of generality, one can choose
\begin{align}
    \chi \in \mathbb{R}, \qquad \eta = |\eta| e^{i \delta}
\end{align}
where the overall phase of the quantum state $\ket{\psi}$ can always be chosen arbitrarily and does not affect measurable quantities such as $g^{(2)}(\psi)$.

With this convention, the normalization condition for the state, as previously derived in Eq.~(\ref{eq:psi_normalization_r=s_f}), now becomes
\begin{align}
\boxed{
|\chi|^2 + |\eta|^2  
+ 2 f(r,\Delta)^{-1/4} |\chi|\, |\eta|\, \cos\big(\delta+\frac{\gamma}{2}\big) = 1
}
\end{align}
where $f(r, \Delta)$ and $\gamma$ are given by Eq.~(\ref{eq:fgamma}).

With this parametrization, all relevant real parts appearing in $g^{(2)}$ can be written explicitly as cosine functions:
\begin{align}
\mathscr{Re}\Big[\chi \eta^* e^{-i \frac{\gamma}{2}}\Big] 
    &= |\chi|\,|\eta| \cos\big(\delta+\frac{\gamma}{2}\big), \nn\\
\mathscr{Re}\Big[\chi \eta^* e^{i(\Delta-\frac{3}{2}\gamma)}\Big] 
    &= |\chi|\,|\eta| \cos\big(\Delta - \frac{3}{2}\gamma - \delta\big), \\
\mathscr{Re}\left[\chi \eta^* e^{i\big(\Delta-\frac52 \gamma\big)}
(2x e^{i\Delta}+1) \right] 
    &= |\chi|\,|\eta| \left[
        2\tanh^2 r \cos\big(2\Delta - \frac{5}{2}\gamma - \delta \big) 
        + \cos\big(\Delta - \frac{5}{2}\gamma - \delta \big) \nn
      \right],
\end{align}
where $x = \tanh^2 r$. Combining all, the closed-form result for $g^{(2)}$ in terms of the phase $\delta$ reads:
\begin{align}
g^{(2)}(\psi) 
=&\Big(3 + \frac{1}{\sinh^2 r}\Big)
\Bigg\{
  1 - 2 f(r, \Delta)^{-1/4} |\chi|\,|\eta| \cos\big(\delta + \frac{\gamma}{2}\big) 
  \label{eq:g2_phase}\\
    &+ 2  \frac{f(r, \Delta)^{-5/4} |\chi|\,|\eta|}{2\tanh^2 r+1}  \Big[
      2\tanh^2 r \cos\big(2\Delta - \frac{5}{2}\gamma - \delta \big)
      + \cos\big(\Delta - \frac{5}{2}\gamma - \delta \big)
    \Big]
\Bigg\} \nn\\
&\times
\Bigg\{
  1 - 2 f(r, \Delta)^{-1/4} |\chi|\,|\eta| \cos\big(\delta + \frac{\gamma}{2}\big)
  + 2 f(r, \Delta)^{-3/4} |\chi|\,|\eta| \cos\big(\Delta - \frac{3}{2}\gamma - \delta\big)
\Bigg\}^{-2}. \nn
\end{align}

\noindent
If either $|\chi| = 0$ or $|\eta| = 0$, then, as expected, one recovers the familiar result for a single squeezed state:
\begin{align}
g^{(2)}(\psi) = 3 + \frac{1}{\sinh^2 r}.
\end{align}
indicating that the nonclassicality (sub-Poissonian statistics) arises exclusively due to interference in the superposition when both components are present.

\subsection{Choosing a specific phase for the superposition} 
Based on the plots, it appears that $\delta=\pi$ and $\Delta=\pm\pi$ minimizes $g^{(2)}$. Let's focus specifically on this case. For this case, we have:
\begin{align}
   \chi = |\chi|,
   \qquad \eta = - |\eta|.
\end{align}
The normalization condition becomes:
\begin{align}
   |\chi|^2 + |\eta|^2
   - 2 \big( 2\sinh^2 r + 1 \big)^{-1/2} |\chi|\,|\eta| = 1.
\end{align}
Also, for $\Delta=\pi$:
\begin{align}
   f(r, \pi) = \big( 2\sinh^2 r + 1 \big)^2 , \quad
   \gamma = 0.
\end{align}
The second-order coherence function $g^{(2)}(\psi)$ is then given by:
\begin{align}
g^{(2)}(\psi)=& \Big(3 + \frac{1}{\sinh^2 r}\Big) \Bigg\{
   1 +2 \big( 2\sinh^2 r + 1 \big)^{-1/2} |\chi|\,|\eta|
  + 2 \big( 2\sinh^2 r + 1 \big)^{-3/2} |\chi|\,|\eta| \Bigg\}^{-2} \nn\\
&
\Bigg\{
   1 +2 \big( 2\sinh^2 r + 1 \big)^{-1/2} |\chi|\,|\eta| 
   + 2 \big( 2\sinh^2 r + 1 \big)^{-5/2} \frac{1 -2\tanh^2 r}{1+2\tanh^2 r}\,  |\chi|\,|\eta|
    \Bigg\}.  \label{eq:g2_pi}
\end{align}

We now have $g^{(2)}$ subject to a constraint. Its minimum can be found using the Lagrange multiplier method. This is a very straightforward problem. For simplicity, let $|\chi|=s$ and $|\eta|=t$, where $s, t > 0$. Thus, we are dealing with three positive variables $r, s, t$, and the expression for $g^{(2)}(\psi)$ becomes:
\begin{align}
g^{(2)}(\psi)=& \Big(3 + \frac{1}{\sinh^2 r}\Big) \Bigg\{
   1 +2 st \big( 2\sinh^2 r + 1 \big)^{-1/2}
  + 2 st \big( 2\sinh^2 r + 1 \big)^{-3/2} \Bigg\}^{-2} \nn\\
&
\Bigg\{
   1 +2 st \big( 2\sinh^2 r + 1 \big)^{-1/2}
   + 2 st  \big( 2\sinh^2 r + 1 \big)^{-5/2} \frac{1 -2\tanh^2 r}{1+2\tanh^2 r}\Bigg\}, \label{eq:g2_pi_2}
\end{align}
subject to the constraint:
\begin{align}
\boxed{  s^2 + t^2
   - 2 st \big( 2\sinh^2 r + 1 \big)^{-1/2} = 1.}
\end{align}

\section{Photon Statistics of the Odd Squeezed Vacuum Superposition}
\label{sec:final_analysis}   
We now present a complete analysis of the second-order correlation function, $g^{(2)}(0)$, for the state engineered to produce strong quantum interference. The state is an ``odd'' superposition of two orthogonally squeezed vacuums, created under conditions of maximal destructive interference (equal amplitudes $|\chi|=|\eta|$ and relative phase $\delta=\pi$). It is defined as:
\begin{equation}
    |\psi\rangle = \mathcal{N} \left( |S(r)\rangle - |S(-r)\rangle \right),
\end{equation}
where $|S(r)\rangle$ is a vacuum state squeezed with a real parameter $r$, $|S(-r)\rangle$ is squeezed along the orthogonal quadrature, and $\mathcal{N}$ is a normalization constant.

\subsection{Fock State Structure and the Minimum Correlation Limit}
To understand the state's properties, we first examine its composition in the photon number basis. The expansion of a general squeezed vacuum state, $|re^{i\theta}\rangle$, contains only even-numbered Fock states $|2n\rangle$. For our two components, we set the squeezing angle to $\theta=0$ for $|S(r)\rangle$ and $\theta=\pi$ for $|S(-r)\rangle$.

The crucial feature of the superposition $|\psi\rangle$ is the quantum interference between these components. The coefficient for each Fock state $|2n\rangle$ is proportional to:
\begin{equation*}
    (-\tanh r)^n - (\tanh r)^n = \begin{cases} 0 & \text{if } n \text{ is even} \\ -2(\tanh r)^n & \text{if } n \text{ is odd} \end{cases}
\end{equation*}
This interference perfectly cancels all Fock state components $|4m\rangle$ (where $n=2m$ is even), including the vacuum state $|0\rangle$. The final, normalized state is therefore a superposition of only the Fock states $|2\rangle, |6\rangle, |10\rangle, \dots$:
\begin{equation}
    |\psi\rangle = \frac{-2\mathcal{N}}{\sqrt{\cosh r}} \sum_{n= \text{odd}}^{\infty} \frac{\sqrt{(2n)!}}{2^n n!} (\tanh r)^n |2n\rangle,
    \label{eq:odd_squeezed_cat_fock_merged}
\end{equation}
where the normalization constant is given by $\mathcal{N} = \left[2\left(1 - (\cosh 2r)^{-1/2}\right)\right]^{-1/2}$.

This structure immediately explains the behavior at zero squeezing. In the limit $r \to 0$, the sum is dominated by its first term ($n=1$), and the entire state converges to the \textbf{two-photon Fock state}, $|\psi\rangle \to |2\rangle$. For the $|2\rangle$ state, the correlation is exactly $g^{(2)}(0)=1/2$, which represents the fundamental minimum for this system.
\subsection{Exact Correlation Function and Asymptotic Behavior}
While the limit at $r=0$ is insightful, the behavior for any value of $r$ can be found from an exact closed-form expression for $g^{(2)}(r)$. Through a lengthy algebraic simplification of the full moment calculation, one arrives at the following rational function of $x = \sinh^2 r$:
\begin{align}
g^{(2)}(r) &= \frac{12 \sinh^{10} r + 40 \sinh^{8} r + 51 \sinh^{6} r + 28 \sinh^{4} r + 11 \sinh^{2} r + 2}{4 \sinh^{10} r + 16 \sinh^{8} r + 29 \sinh^{6} r + 29 \sinh^{4} r + 16 \sinh^{2} r + 4} \label{eq:g2_exact_merged} \\
&\approx \frac{1}{2} + \frac{3}{4} \sinh^2 r + \frac{3}{8} \sinh^4 r 
+ \frac{35}{16} \sinh^6 r + O(\sinh^8 r).
\end{align}
This exact formula allows us to analyze the state's properties across all squeezing regimes. The Taylor series confirms that the function starts at 1/2 and initially increases with $r$.

To understand the behavior for large squeezing, we analyze the asymptote of Eq.~\eqref{eq:g2_exact_merged} as $r \to \infty$. The limit is given by the ratio of the leading-order terms:
\begin{equation}
    \lim_{r\to\infty} g^{(2)}(r) = \lim_{x\to\infty} \frac{12x^5 + \dots}{4x^5 + \dots} = \frac{12}{4} = 3.
\end{equation}
This asymptotic value of 3 is the well-known limit for a single, highly squeezed vacuum state. It indicates that for very large squeezing, the interference effects become negligible, and the photon statistics are dominated by the properties of the individual squeezed components.

In summary, the analysis shows that $g^{(2)}(0)$ for this state has a global minimum of 1/2 at $r=0$ and monotonically increases towards an asymptote of 3 as the squeezing grows.

\subsection{\texorpdfstring{The derivation of the boundary value \( g^{(2)}_{\text{boundary}}(r) \)}{}}

\begin{table}[t]
\caption{Calculated $g^{(2)}(\psi)$ values for varying squeezing parameter $r=s$, with fixed optimal relative phase $\Delta=\pi$, relative amplitude phase $\delta=\pi$, and superposition amplitude $|\eta|=2.20070$, illustrating the local minimum.}
\label{tab:g2_r_scan_optimal_eta}
\centering
\begin{tabular*}{\linewidth}{@{\extracolsep{\fill}}ccc}
\hline\hline\noalign{\smallskip}
$r$ & $|\eta|$ & $g^{(2)}(\psi)$ \\
\hline\noalign{\smallskip}
0.26 & 2.20070 & 0.58418 \\
0.28 & 2.20070 & 0.57467 \\
0.30 & 2.20070 & 0.56942 \\
0.32 & 2.20070 & 0.56770 \\
0.34 & 2.20070 & 0.56740 \\
0.36 & 2.20070 & 0.56930 \\
0.38 & 2.20070 & 0.57245 \\
0.40 & 2.20070 & 0.57723 \\
\noalign{\smallskip}\hline\hline
\end{tabular*}
\end{table}

The derivation of the boundary value \( g^{(2)}_{\text{boundary}}(r) \) for the minimum second-order coherence in the symmetric case \( r = s \) is as follows:

\begin{enumerate}
\item For the optimal phases \(\delta = \pi\) and \(\Delta = \pi\), the expression for \( g^{(2)}(0) \) reduces to
\[
 g^{(2)}(\psi) = \left(3 + \frac{1}{\sinh^2 r}\right) \frac{1 + L + L K^4 (1 - 2\tanh^2 r)}{(1 + L + L K^2)^2},
\]
where \( K = (1 + 2 \sinh^2 r)^{-1/2} \) and \( L = 2 K |\chi| |\eta| \), subject to the normalization \( |\chi|^2 + |\eta|^2 - L = 1 \).

\item The minimum occurs at the maximum \(L = K / (1 - K)\).

\item Substituting \(L = K / (1 - K)\) into the expression and simplifying yields the boundary
\[
 g^{(2)}_{\text{boundary}}(r) = \frac{12\sinh^{10} r + 40\sinh^{8} r + 51\sinh^{6} r + 28\sinh^{4} r + 11\sinh^{2} r + 2}{4\sinh^{10} r + 16\sinh^{8} r + 29\sinh^{6} r + 29\sinh^{4} r + 16\sinh^{2} r + 4}.
\]
\end{enumerate}

To arrive at this form, substitute the definitions of \( K \) and \(\tanh r\) in terms of \(\sinh r\), perform algebraic simplification by clearing denominators, and collect like terms in powers of \(\sinh^2 r\). This boundary approaches \(\frac{1}{2}\) as \( r \to 0 \). 

\subsection*{Benchmark values near the practical minimum}
For convenience, Table~\ref{tab:g2_r_scan_optimal_eta} lists representative values
along the optimal ridge ($r=s$, $\Delta=\pi$, $\delta=\pi$) around the local
minimum ($r\approx 0.34$).

\bibliography{SqueezingRef}


\end{document}